\documentclass[prl,showpacs,preprint, preprintnumbers]{revtex4}%
\usepackage{amsfonts}
\usepackage{amsmath}
\usepackage{amssymb}
\usepackage{graphicx}%
\providecommand{\U}[1]{\protect\rule{.1in}{.1in}}

\begin{document}
\title{A semi-quantitative scattering theory of amorphous materials}
\author{Mingliang Zhang}
\email{zhang@phy.ohiou.edu}
\author{Yue Pan}
\email{pan@phy.ohiou.edu}
\author{F. Inam}
\email{inam@phy.ohiou.edu}
\author{D. A. Drabold}
\email{drabold@ohio.edu}
\affiliation{Department of Physics and Astronomy, Ohio University, Athens, OH 45701}

\begin{abstract}
It is argued that topological disorder in amorphous solids can be described by
local strains related to local reference crystals and local rotations. An
intuitive localization criterion is formulated from this point of view. The
Inverse Participation Ratio and the location of mobility edges in band tails
is directly related to the character of the disorder potential in amorphous
solid, the coordination number, the transition integral and the nodes of wave
functions of the corresponding reference crystal. The dependence of the decay
rate of band tails on temperature and static disorder are derived. \textit{Ab
initio} simulations on a-Si and experiments on a-Si:H are compared to these predictions.

\end{abstract}

\maketitle

Electronic localization induced by diagonal disorder or by structural disorder
has been intensively studied over nearly fifty years\cite{kra}. However, key
properties like the energy dependence of the Inverse Participation Ratio
(IPR), the location of the mobility edges and the  decay rate of and tails are
expressed in an obscure way, not directly accessible to experiment or
simulation\cite{edg}. Perturbation theory has been applied to approximate the
electron states of amorphous solids (AS), starting with a crystalline
counterpart as zero order solution\cite{mofe} even before Anderson's classical
work\cite{edg}. In this Letter we suggest that a local formulation of
perturbation theory is effective for the localized states confined to one
distorted region and for the first time relate important physical quantities
such as the decay rate of band tails and energy dependence of IPR etc. to
basic material properties.

Similar to the theory of elasticity\cite{lov}, the distorted regions in AS can
be characterized by local strains referring to their local reference crystal
(LRC) and local rotations. By a suitable choice of origin and orientation of
LRC, the atomic displacements of a distorted region of AS relative to its LRC
are small. Thus the relative change in potential energy for each distorted
region in AS is small. Perturbation theory is justified for each distorted
region. The semi-classical approximation (SCA)\cite{pipa} can further simplify
the calculation of scattering waves caused by a distorted region, since the de
Broglie wavelength for low-lying excitations is of order one bond length
($\approx$2.35 \AA \ in a-Si\cite{har}), a distance much shorter than the
characteristic range in which the random potential fluctuates\cite{ori,yp1}.
The motion of electronic packet under extra force of AS relative to LRC can be
described by the Ehrenfest theorem\cite{pipa}.

We first formulate an intuitive localization criterion for the states confined
to one distorted region. Then the IPR, the position of mobility edge and
Urbach energy are related to the distortion relative to the LRC, the
coordination number and the inter-cell transition integral. The predictions
are consistent with available experiments. We also performed \textit{ab
initio} local density approximation (LDA)\cite{sie} and tight binding
approximation (TBA)\cite{dra,lud} computations on a-Si to verify our results.

Consider a distorted region $\mathcal{D}$, with linear size $L$. Using the
primitive cell of LRC numbering the atoms in $\mathcal{D}$, the x-component of
extra force suffered by an electron relative to that of LRC is%
\begin{equation}
F_{x}(\mathbf{r})=\sum_{\mathbf{n}\beta}\frac{\partial^{2}U(\mathbf{r}%
-\mathbf{R}_{\mathbf{n}})}{\partial R_{\mathbf{n}x}\partial R_{\mathbf{n}%
\beta}}u_{\mathbf{n}\beta}^{s},\text{ \ \ \ }\beta=x,y,z \label{for}%
\end{equation}
later its characteristic value of is denoted as $F$. $\mathbf{n}$ is lattice
index, $\mathbf{R}_{\mathbf{n}}$ and $u_{\mathbf{n}\beta}^{s}$ are the
position vector and the $\beta$th component of the static displacement of the
atom $\mathbf{n}$ respectively. $U(\mathbf{r}-\mathbf{R}_{\mathbf{n}})$ is the
potential energy felt by an electron at $\mathbf{r}$ from the atom at
$\mathbf{R}_{\mathbf{n}}$.

A Bloch wave $\psi_{n\mathbf{k}}^{c}$ of LRC passes through $\mathcal{D}$, and
in SCA\cite{pipa}, the change in the $x$ component of the wave vector after
scattering is
\begin{equation}
\Delta k_{x}\thicksim\frac{(FL)_{x}}{\nabla_{k_{x}}E_{n\mathbf{k}}} \label{cp}%
\end{equation}
$FL$ measures the magnitude of random potential in $\mathcal{D}$. The phase
shift $\delta_{n\mathbf{k}}$ of state $\psi_{n\mathbf{k}}^{c}$ \bigskip is
determined by the change in momentum and the propagation path of the Bloch
wave%
\begin{equation}
\delta_{n\mathbf{k}}\thicksim\frac{FL^{2}}{|\nabla_{\mathbf{k}}E_{n\mathbf{k}%
}|} \label{ps}%
\end{equation}
where $E_{n\mathbf{k}}$ is dispersion relation of the $n$th energy band of the
LRC. $FL^{2}$ is the strength of a potential well (the product of the depth of
potential well and the range of force) in standard scattering theory\cite{cal}%
. If the first coordination shell around an atom is spherically symmetric, the
dispersion relation in TBA is\cite{cal}
\begin{equation}
E_{n\mathbf{k}}\thicksim E_{n0}-zI_{n}\cos k_{x}a\text{ .} \label{ds}%
\end{equation}
Here $E_{n0}$ is the middle of the $n$th band ($k_{x}a=\pi/2$), $z$ is the
coordination number of a cell, $I_{n}$ is the transition integral for the
$n$th band, $a$ is the lattice constant in LRC. For a semi-quantitative
discussion, crude dispersion relation (\ref{ds}) will not invalidate essential
points. If the phase shift $\delta_{n\mathbf{k}}$ of the secondary scattering
waves relative to the primary wave is $\thicksim\pi$, then outside
$\mathcal{D}$, scattering waves will interfere destructively with the primary
Bloch state. No probability amplitude appears outside $\mathcal{D}$. A
localized state is therefore formed inside $\mathcal{D}$ due to the
constructive interference of a Bloch state $\psi_{n\mathbf{k}}^{c}$ and its
secondary scattering waves.

\begin{figure}[h]
\begin{center}
\resizebox{65mm}{!}{\includegraphics{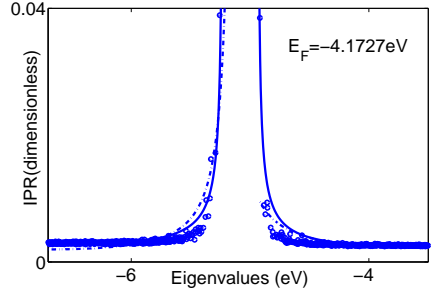}}
\end{center}
\caption{IPR of 512-atom model of a-Si, dots from \textit{ab initio}
calculation\cite{sie}, dashed line and solid line are from two parameter ($FL$
and $zI$) least squares fit and eye guide fit with Eq.(\ref{ipre}).}%
\label{Fig1}%
\end{figure}

Bloch states of LRC at top of valence and at bottom of the conduction edges
are susceptible to the random potential. The former is shorter wave, sensitive
to details of atomic displacements of a distorted region. The latter is long
wave: a small random potential will easily produce a change in momentum
comparable to $\hbar\mathbf{k}$ itself. In other words, states with small
group velocity are easily localized. The group velocity of an electron in
state $\psi_{n\mathbf{k}}^{c}$ in TBA is $v_{n\mathbf{k}}^{g}\thicksim
\frac{zI_{n}a}{\hbar}\sin k_{x}a$, states near to bottom ($k_{x}a\thicksim0$)
and states near to top ($k_{x}a\thicksim\pi$ ) have small $v_{n\mathbf{k}}%
^{g}$. According to Eq.(\ref{ps}), they are more easily localized than the
states in the middle of a band for a given random potential. For $k$ close to
$\frac{\pi}{a}$, with TBA dispersion relation (\ref{ds}), group velocity of
state $\psi_{k}^{v}$ is $v_{k}^{g}=\frac{Iz}{\hbar}(\frac{E_{0}-E}{Iz})^{1/2}%
$, $E_{0}=E_{0}^{V}+zI_{V}$ is the top of the valence band. By Eq. (\ref{ps}),
under TBA, for a valence state $\psi_{k}^{v}$ with energy $E_{k}$, the change
in phase shift with energy is $\frac{d\delta_{k}}{dE}=\frac{FL^{2}}%
{a(E_{0}-E_{k})^{3/2}(Iz)^{1/2}}$. For a given distorted region, Bloch states
close to $E_{0}$ will suffer larger phase shift. They are more readily
localized than the states in the middle of the band. Similar conclusion holds
for the Bloch states in the bottom of conduction band. In Fig. \ref{Fig1} the
IPR is plotted against electron energy for a model of a-Si. Large IPR appears
at the edges of a band, in agreement with the above prediction.

The upper mobility edge of the valence band is the deepest energy level
$E_{\mathbf{k}_{\ast}^{V}}^{V}$that the largest distorted region could
localize, i.e. produce a phase shift $\pi$ for the corresponding Bloch state.
In TBA, this leads to $\sin k_{\ast}^{V}a=\frac{FL^{2}}{zI^{V}a\pi}$. The
energy difference between the top of a band and the mobility edge is
$E_{me}^{V}=z_{V}I^{V}\{1-[1-(\frac{FL^{2}}{z_{V}I^{V}a\pi})^{2}%
]^{1/2}\}\thicksim\frac{(\frac{FL^{2}}{a\pi})^{2}}{z_{V}I^{V}}$, last
$\thicksim$ only holds for $\frac{FL^{2}}{z_{V}I^{V}a\pi}<<1$. It is obvious
that stronger random potential and narrower band lead to a deeper mobility
edge. The lower mobility edge of the conduction band can be obtained
similarly. The energy difference $\Delta_{m}$ between the lower mobility edge
of the conduction band and the upper edge of the valence band is%
\begin{equation}
\Delta_{m}\thickapprox G^{C}+[\frac{(\frac{FL^{2}}{a\pi})^{2}}{z_{V}I^{V}%
}+\frac{(\frac{F_{C}L_{C}^{2}}{a_{C}\pi})^{2}}{z_{C}I^{C}}] \label{edf}%
\end{equation}
where $G^{C}$ is the band gap of LRC. Because the van Hove singularity is
smeared out in AS, gap in amorphous solid is ambiguous. $\Delta_{m}$ can be
defined in a simulation by identifying two edge states.

In the middle of a band $k_{x}a=\frac{\pi}{2}$, the group velocity reaches its
maximum $\frac{zI_{n}a}{\hbar}$. By Eq. (\ref{ps}), to localize the states in
the middle of the $n$th band, we need $\frac{FL}{zI_{n}}\frac{L}{a}\gtrsim\pi
$. States in the middle of a band are most difficult to localize. If those
states are localized, the whole band is localized. A stronger localization
condition is $\Delta k\thicksim k$. In the middle of band $k_{x}=\frac{\pi}%
{2}\frac{1}{a}$, by Eq. (\ref{cp}) the change in wave vector is $\frac
{FL}{zI_{n}a}$. It leads to the condition to localize a whole band $\frac
{FL}{zI_{n}}\gtrsim\frac{\pi}{2}$ (smaller than $\frac{FL}{zI_{n}}%
\thicksim6-34$)\cite{sri}. The deeper localized states in AS are generated by
the deeper Bloch states of LRC, are spread in several distorted regions.
Because current local description only considers the states localized in one
distorted region, we cannot expect a better estimate.

The IPR $I_{j}$ of a localized eigenstate $\psi_{j}$ could be approximated
as\cite{kra} \ $I_{j}\thicksim\frac{a^{3}}{\xi_{j}^{3}}$, $\xi_{j}$ is the
localization length of $\psi_{j}$. If a Bloch wave $\psi_{n\mathbf{k}}^{c}$
suffers a phase shift $\pi$ by some distorted region to produce $\psi_{j}$, it
is localized in range $\xi_{j}:$ $\xi_{j}\Delta k\thicksim\pi$. The change in
wave vector is $\Delta\mathbf{k}\thicksim\frac{FL}{\nabla_{\mathbf{k}%
}E_{\mathbf{k}}}$,%
\begin{equation}
\xi_{j}\thicksim\frac{\pi}{\Delta k}=\frac{\pi\nabla_{k}E_{k}}{FL}%
\thicksim\frac{\pi zI_{n}a\sin ka}{FL} \label{ll}%
\end{equation}
($\thicksim$ is obtained under TBA). According to Eq.(\ref{for}),
$F\thicksim\epsilon$, $\epsilon$ is the relative change in lattice constant.
To minimize the free energy, a denser region with shorter bonds and small
angles will gradually decay away toward the mean density rather than exhibit
an abrupt transit to a diluter region and \textit{vice versa}. Therefore the
size $L$ of a denser distorted region is proportional to $\epsilon$. Eq.
(\ref{ll}) indicates $\xi\thicksim\frac{a}{\epsilon^{2}}$\cite{mofe}. The
advantage of Eq.(\ref{ll}) is that it reveals the role of the coordination
number $z$ and the transition integral $I$. The dependence on $\mathbf{k}$
(wave length and propagation direction of Bloch wave) is also displayed in Eq.
(\ref{ll}): close to band edge of LRC, $ka\thicksim0$ or $\frac{\pi}{a}$, the
localization length is small and IPR is high (see Fig. \ref{Fig1}).

Making use of Eqs. (\ref{ll}) and (\ref{ds}),%
\begin{equation}
\xi_{j}(E_{k^{j}})=\frac{\pi zI_{V}a}{FL}[1-(\frac{E_{k^{j}}-b_{me}^{V}%
+zI_{V}-E_{me}^{V}}{I_{V}z})^{2}]^{1/2} \label{leno}%
\end{equation}
$b_{me}^{V}$ is the location of the mobility edge of valence band. When we
approach $b_{me}^{V}$ from the upper side with higher energy, it is easy to
find $\xi_{j}\rightarrow L$ from Eq. (\ref{leno}), localization length $\xi$
approach to the size $L$ of whole sample as $(E_{k^{j}}-b_{me}^{V})^{\alpha},$
where $\frac{1}{2}<\alpha<1,$ it is close the lower bound of previous
works\cite{mcon}. The trend expressed by (\ref{leno}) is consistent with a
simulation based upon time-dependent Schrodinger equation\cite{rae}.

For a localized state derived from Bloch wave $\psi_{k^{j}}^{c}$ in LRC, the
energy dependence of IPR can be found%
\begin{equation}
I(E_{k^{j}})\thicksim\frac{(FL/\pi zI_{V})^{3}}{[1-(\frac{E_{k^{j}}-E_{0}^{V}%
}{zI_{V}})^{2}]^{3/2}} \label{ipre}%
\end{equation}
This is a new prediction of our work. Eqs. (\ref{leno}) and (\ref{ipre}) are
not quite satisfied because $E_{k^{j}}$ is the corresponding energy level in
LRC, not the eigenvalue of the localized state $\psi_{j}^{a}$. It can be cured
by taking into account energy level shift caused by the disorder in AS
relative to LRC. Fig. \ref{Fig1} shows IPR vs. eigenvalues in a 512-atom model
of a-Si\cite{ori}. As expected from Eq. (\ref{ipre}), IPR decreases from
highest values at band edges to small values in the band interior. The
functional form (\ref{ipre}) fits the simulation rather well.

According to Eq. (\ref{ipre}), the least squares fitting
parameters in Fig. \ref{Fig1} are $(FL)_{V}=1.256$eV, $(zI)_{V}=3.185$eV,
$E_{0}^{V}=-7.390$eV$,$ $(FL)_{C}=1.437$eV, $(zI)_{C}=3.502$eV, $E_{0}%
^{C}=-1.080$eV. The width of valence band of c-Si is about 2.7eV, the width of
conduction band is about 2.3eV\cite{col}. The fit parameters are
reasonable-something like what we expect for Si. Gap for c-Si is
1.12eV\cite{col}, using above parameters with help of Eq.(\ref{edf}), the
distance between mobility edges is 2.205eV. Result from LRC model falls in the
range 1.58-2.43 eV of the observed optical gap\cite{den,kod,hag}.

In a distorted region of a-Si where bonds are shortened, valence states have
more amplitude in the middle of bonds. Random potential $V_{a}-V_{c}$ (the
difference between the amorphous and crystalline potentials) is important only
in the middle of bonds rather than close to the core of atoms. Electrons will
feel $V_{a}-V_{c}$ more than a region where bonds more close normal. Valence
tail states are easily localized in a distorted region with shorter
bonds\cite{ori,jj}. On the other hand, in a distorted region with longer
bonds, the conduction levels are lowered and the probability of conduction
electrons staying in the middle of nearest neighbor atoms becomes larger than
a region where bonds are closer to the mean. Conduction tail states are more
readily localized in a distorted region with longer bonds and large
angles\cite{ori,jj}.

The effect of three- and four- point correlation on the shape of band tail is
subtle: localized states adhere to 1D filaments in AS network\cite{yp1}. In
the spirit of scattering theory of line shape\cite{sto}, the decay rate
$E_{U}^{V(C)}$ of valence (conduction) tail can be derived from the relative
shift of energy levels of LRC. Suppose $\Delta b$ is the distribution width of
bond length (BL), the blurring $\delta k$ in wave vector $k$ is $\frac{\Delta
b}{b}k$. The shift of level $E_{k}^{v}$ ($E_{k}^{c}$) for a Bloch state
$\psi_{k}^{v}$ ($\psi_{k}^{c}$) in valence (conduction) band by the disorder
in AS is $\Delta E_{k}^{v(c)}=\int d\tau(V_{a}-V_{c})|\psi_{k}^{v(c)}|^{2}$.
The relative level shift due to this BL distribution is $\delta k\frac{d}%
{dk}\Delta E_{k}^{c}$. It is easy to see $V_{a}-V_{c}\varpropto\frac{\Delta
a}{a}V_{c}$. Then%
\begin{equation}
E_{U}^{V(C)}\thicksim\frac{\Delta b}{b}k\cdot\frac{1}{k}\frac{\Delta b}%
{b}V_{c}=(\frac{\Delta b}{b})^{2}|V_{c}|=\frac{(\frac{\Delta b}{b}|V_{c}%
|)^{2}}{|V_{c}|}\label{urc}%
\end{equation}
If we make a correspondence between structural disorder $\frac{\Delta b}%
{b}|V_{c}|$ and on-site spread $W$ of levels, Eq.(\ref{urc}) is comparable to
$E_{U}\thicksim0.5\frac{W^{2}}{B}$ ($B$ is the band width) \cite{at} and
$E_{U}\thicksim\frac{\pi}{4}\frac{W^{2}}{3\pi^{2}\frac{\hbar^{2}}{2mL^{2}}}$
\cite{sce}, where $L$ and $W$ are correlation length and variance of random
potential. Eq. (\ref{urc}) is also consistent with an assumption of Cody et.
al. to explain their absorption edge data in a-Si:H\cite{cod}. Since the width
of BL distribution is $\frac{\Delta b}{b}\thickapprox0.1$ and $|V_{c}%
|\thicksim1-10$eV, the order of magnitude of mobility edge should be
$(\frac{\Delta b}{b})|V_{c}|$, several tenth eV to 1eV, so that the decay rate
of band tails is around several tens to several hundred meV. Both agree with
experimental observations\cite{sher}. Eq.(\ref{urc}) indicates $E_{U}^{V(C)}$
is proportional to static disorder that is characterized by $(\frac{\Delta
b}{b})^{2}$, in consistent with the fact that $E_{U}^{V(C)}$ of a-Si:H
increased with deposition power\cite{sher}. $\Delta b$ and $b$ could also be
explained as the width and the average value of BA distribution.

Because local compression is compensated by adjacent local tensile in AS,
$E_{U}^{V}\thicksim\frac{\varsigma^{V}}{\varsigma^{C}}E_{U}^{C}$, where
$\varsigma^{V}$ ($\varsigma^{C}$) is an order one dimensionless constant
characterizing the peak (node) of valence (conduction) states. In a-Si and
a-Si:H, random potential $(V_{a}-V_{c})$ has larger distortion in the middle
of Si-Si bonds, since valence states are more in the middle of bonds than
conduction states\cite{har}, they feel the distortion more. Therefore
$\varsigma^{V}>\varsigma^{C}$. One expect $E_{U}^{V}$ $>$ $E_{U}^{C}$. This
agrees with measurements in a-Si:H: $E_{U}^{V}\thicksim$43-103meV vs.
$E_{U}^{C}\thicksim$27-37 meV, linear relation among $E_{U}^{V}$ and
$E_{U}^{C}$ has also been observed\cite{sher}.

\begin{figure}[h]
\begin{center}
\resizebox{80mm}{!}{\includegraphics{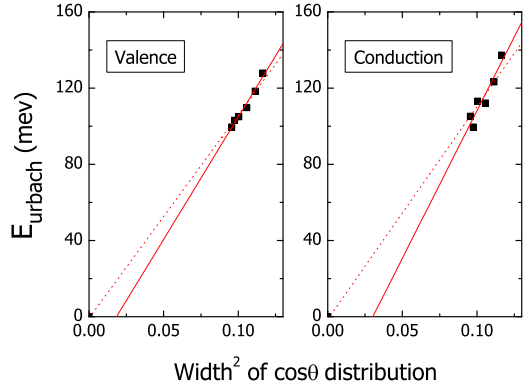}}
\end{center}
\caption{$E_{U}^{V}$ and $E_{U}^{C}$ vs. $\sigma_{cos\theta}^{2}$: 6 squares
are extracted from TBA, dotted line and solid line are least square fits with
and without (0,0) points.}%
\label{Fig2}%
\end{figure}To test correctness of Eq.(\ref{urc}), we undertook a TBA
calculation for DOS of six a-Si models with 20,000 atoms\cite{dra,lud,mou}.
$E_{U}^{V(C)}$, the width $\sigma_{\cos\theta}$ of BA distributions and the
width $\Delta b$ of BL distribution are extracted. Fig.\ref{Fig2} clearly
shows good linear relation between $E_{U}^{V}$ ($E_{U}^{C}$) and $\sigma
_{\cos\theta}^{2}$, curves pass origin ($E_{U}^{V(C)}$ is zero for crystal) as
displayed in Eq. (\ref{urc}). It can be further tested in ion implanted
samples, where a continuous increase disorder from crystal to amorphous are
realized by increasing the dose\cite{sor}. The $E_{U}^{V}$ ($E_{U}^{C}$) vs.
($\Delta b)^{2}$ curve does not pass origin (not showing here), this is an
indication that BA disorder is a little more decisive in determine the shape
of a band tail than BL disorder for a well relaxed structure\cite{ori,jj}.

The electron-phonon interaction is strong in AS\cite{atf}. At finite
temperature, the displacement of an atom in AS deviate from the position in
the LRC at zero temperature is a vector sum of the static displacement
$\mathbf{u}_{s}$ and thermal vibration displacement $\mathbf{u}_{T}(t)$ from
the zero temperature configuration of AS, $t$ is the time moment. In ordinary
absorption experiment, time interval $T$ is much longer than the period of the
slowest mode, therefore $\overline{E_{U}^{C}}=\frac{1}{T}\int_{0}%
^{T}dt\varsigma^{C}(\frac{\mathbf{u}_{s}+\mathbf{u}_{T}(t)}{a})^{2}|V_{c}|$.
Atoms vibrate around their equilibrium points in AS, the time average of the
cross term $\mathbf{u}_{s}\cdot\mathbf{u}_{T}(t)$ is zero. Thus Urbach energy
from static disorder and from thermal disorder is additive\cite{cod}
$\overline{E_{U}^{C}}=E_{Us}^{C}+E_{UT}^{C},$ $E_{Us}^{C}=\varsigma^{C}%
(\frac{u_{s}}{a})^{2}|V_{c}|$.\ Thermal part $E_{UT}^{C}=\varsigma^{C}%
\frac{\overline{\mathbf{u}_{T}^{2}}}{a^{2}}|V_{c}|$, $\overline{\mathbf{u}%
_{T}^{2}}=\frac{1}{T}\int_{0}^{T}dt[\mathbf{u}_{T}(t)]^{2}$ is the long time
average of the square of amplitude of vibration. An ultra-fast probe of
absorption edge may find oscillating in $\overline{E_{U}^{C}}$. Since
$\overline{\mathbf{u}_{T}^{2}}\varpropto\frac{k_{B}T}{B_{C}}a^{2}$\cite{cal},
$B_{C}$ is binding energy in the diluter regions where conduction tail states
are localized, $E_{UT}^{C}=\varsigma^{C}k_{B}T\frac{|V_{c}|}{B_{C}}$.
$\overline{E_{U}^{C}}$ linearly increases with temperature. Similarly result
holds for $\overline{E_{U}^{V}}$. The is consistent with the fact that above
350K absorption edge linearly increase with $k_{B}T$ in a-Si:H\cite{wei,alj}.
Because $B_{V}>B_{C}$, $E_{U}^{C}$ is more susceptible to thermal disorder
than $E_{U}^{V}$\cite{dra1}, as observed in ref. \cite{alj}.

For \textit{realistic} amorphous solid with \textit{topological disorder}, by
viewing an AS as many distorted regions relative to corresponding LRC, we push
forward essential understanding on localized states confined in one distorted
region. The predicted IPR, mobility edge, the dependence on static disorder
and on temperature of the decay rate of band tails agree with available
experiments and simulations. We explained the fact that valence tail states
are more localized in a denser region with smaller BA and shorter BL and
conduction tail states are more localized in diluter region with longer BL and
larger BA in a-Si\cite{ori,jj}. Localized states in several distorted regions
and other problems involving global topology will be addressed in future.

{\large Acknowledgements:} We thank the Army Research Office for support under
MURI W91NF-06-2-0026, and the National Science Foundation for support under
grant No. DMR 0600073, 0605890.

\end{document}